\documentclass[a4paper]{jpconf}
\usepackage{graphicx}
\usepackage{enumerate}
\usepackage{amssymb, amsmath, amsfonts, amsthm}
\usepackage[utf8]{inputenc} 
\usepackage{float}
\usepackage{graphicx}
\usepackage{flushend}
\usepackage{multicol}

\usepackage{multirow}
\usepackage{slashed}
\usepackage{mathrsfs}
\newcommand{\s}{\slashed}
\newcommand{\g}{\gamma}
\newcommand{\ba}{\begin{align}}
\newcommand{\ea}{\end{align}}
\newcommand{\be}{\begin{equation}}
\newcommand{\ee}{\end{equation}}

\begin{document}

\title{Helicity Amplitudes for production of massive gravitino/goldstino}

\author{Bryan O. Larios}

\address{Facultad de Ciencias F\'isico - Matem\'aticas, BUAP \protect\\Apdo. Postal 1364, C.P. 72000, Puebla, Pue. M\'exico}

\ead{bryanlarios@gmail.com}
%%%%%%%%%%%%%%%%%%%%%%%%%%%%
\begin{abstract}
The spinor helicity formalism (SHF) has been efficiently  applied to compute  perturbative  amplitudes in plenty of processes and reactions in gauge theories (including gravity), mostly in the massless case. Some work has been done in order to extend the SHF to the massive case. We have used these powerful tools to evaluate amplitudes in a local supersymmetric model where the gravitino is the lightest supersymmetric particle (LSP).  Two decays have been evaluated in order to show the capability of the SHF in the massive extension, namely the two body neutralino decays $\tilde{\chi}_0\to V\tilde{\Psi}^{\mu}$ with $V=\gamma, Z$. The comparisons of amplitudes with spin-3/2 gravitino and spin-1/2 goldstino are also presented.
\end{abstract}
%%%%%%%%%% Introduction %%%%%%%%%
\section{Introduction}
%%%%%%%%%%%%%%%%%%%%%%%%%%%%%%%%%%%%%%%%%%
The traditional Feynman approach for perturbation theory has been developed for almost sixty years, and it has been undoubtedly  a magnificent tool to evaluate amplitudes in gauge theories. Even with all the success of the Feynman rules, several computations for scattering processes especially in QCD suffer the lack of accuracy in order to be tested experimentally. Furthermore, processes with several external particles are extremely complicated to calculate with the traditional perturbative approach \cite{dixon}.  Recent developments for tree-level scattering  amplitudes in the massless case using the helicity methods are now widely used \cite{parke, srednicki,  peskin2, schwartz, henn, elvang}, allowing to efficiently evaluate  amplitudes for processes with $n$ external legs in pure Yang-Mills theories. A huge theoretical progress has been done  during the last two decades in this active and growing field, making manifest certain mathematical structures that underlie quantum field theories \cite{witten, bcfw, mazon, carrasco, nima5, cachazo}.\newline  
\indent We shall try to present a pragmatic approach in this work, starting still \textit{\`a la Feynman},  building the scattering amplitudes from the Feynman diagrams but expressing the usual spinors (in any representation)  appearing in the amplitudes as momentum twistors, this step shall hugely facilitate the calculations, as we shall see later. This is one of the goals of the SHF. Addressing the massive case requires the implementation of the so-called light cone decomposition (LCD) for the massive momenta \cite{ozeren, spinorsextras, kosower, weinzierl}. This technique allow us to express a massive momentum, namely $p^2=-m^2$ as a lineal combination of two massless momentum $p_i=r_i+\alpha q_i$ where $r_i^2=q_i^2=0$ and $\alpha=-\frac{m^2}{2q_i\cdot r_i}$.  Several processes of the electroweak Standard Model have been evaluated  using the massive SHF in Ref. \cite{bryan-oscar}, reproducing the well known results. More recently we have implemented this method for $\mathcal{N}=1$ supergravity with LSP gravitinos \cite{lorenzo-bryan}, showing that the formalism is well suitable for amplitudes involving massive gravitinos in the final state. 
\newline 
\indent Considering scenarios where the gravitino is the LSP and hence a good dark matter candidate,  the nature of the next-to-lightest  supersymmetric particle (NLSP) determines its phenomenology \cite{feng1, feng2, ellis1, ellis2,lorenzo-bryan2}. Potential candidates for NLSP include  the lightest neutralino \cite{steffen, johansen}, the chargino \cite{kribs}, and the lightest charged slepton \cite{heisig}. The weakness of the gravitational interactions suggest that the NLSP could have a long lifetime leading to scenarios with a metastable charged sparticles that could have striking signatures at colliders \cite{ellis3, kuno} and it could also affect the Big Bang nucleosynthesis  \cite{ellis5, kohri, kohri2}. Here we shall consider several decay modes for the  neutralino as the NLSP.  We also have considered the case 
when the gravitino can be approximated
by the goldstino state, this is for the region where the gravitino mass is small compared to the neutralino mass $\tilde{m}\ll m_{\tilde{\chi}_0}$.
This proceeding contribution is structured as follows: Section \ref{section2} contains the solutions of the Rarita-Schwhinger equation for the gravitino. The four states of the  gravitino  are expressed in terms of the momentum twistors after LCD  is applied. We shall see that this is the key to obtain compact expression for the decay amplitudes.  In Section \ref{section3} the helicity amplitudes (HAs) for the two decays ($\tilde{\chi}_0\to V\tilde{\Psi}^{\mu}$ with $V=\gamma, Z$) are shown in terms of very compact mathematical expressions, in addition the HAs with the golsdtino approximation are also shown. Finally  in Section \ref{conclussions} we present some final remarks and comments.
%%%%%%%%%%%%%%%%%%%%%%%%%%%%%%%%%%%%
\section{Gravitino wave functions}\label{section2}
\subsection{Rarita-Schwinger equations}
\indent The wave function for massive gravitino is the solution of the Rarita-Schwhinger equation \cite{rarita}, this equation of motion (EOM) results from applying the Euler-Lagrange to the following lagrangian \cite{takeo}
\begin{equation}\label{lagrangian}
\mathcal{L}=-\frac{1}{2}\epsilon^{\mu\nu\rho\sigma}\Psi_\mu^TC^{\dagger}\gamma_5\gamma_\nu\partial_\rho\Psi_\sigma+\frac{1}{4}\tilde{m}\Psi_{\mu}^TC^{\dagger}[\gamma^{\mu},\gamma^{\nu}]\Psi_{\nu},
\end{equation}
the  EOM for the gravitino  are equivalent to the following equations \cite{auvil}
\begin{align}\label{eq:01}
\gamma^{\mu}\Psi_{\mu}&=0,\\\label{eq:02}
\partial^{\mu}\Psi_{\mu}&=0,\\\label{eq:03}
(i\gamma^{\mu}\partial_{\mu}-\tilde{m})\Psi_{\mu}&=0,
\end{align}
where $\tilde{m}$ is the gravitino mass. Through all this work we shall use the following convention for the Minkowski metric $\eta^{\mu\nu}=\text{diag}(-1,1,1,1)$, besides we have used the  Dirac representation for the gamma matrices where they take the following form
\begin{equation}\label{diracgamma}
\g^{\mu}=\left(\begin{array}{c c }
		0 & \sigma^{\mu} \\ 
		\bar{\sigma}^{\mu} & 0
	\end{array}\right),
\end{equation}
with $\sigma^{\mu}=(1,\vec{\sigma})$ and $\bar{\sigma}^{\mu}=(1,-\vec{\sigma})$. Returning to the EOM Eqs.~(\ref{eq:01})-(\ref{eq:03}), these admit the following solution \cite{auvil} 
\begin{equation}\label{CG:equation}
\tilde{\Psi}_{\mu}(\vec{p},\lambda)=\sum_{s,m}\left\langle\left(\frac{1}{2},\frac{s}{2}\right)(1,m)\Bigg|\left(\frac{3}{2},\lambda\right)\right\rangle u(\vec{p},s)\epsilon_{\mu}(\vec{p},m),
\end{equation}
this is a Clebsch-Gordon expansion with coefficient $\langle\left(\frac{1}{2},\frac{s}{2}\right)(1,m)\Big|\left(\frac{3}{2},\lambda\right)\rangle$, from Eq.~(\ref{CG:equation}), we know that there are four gravitino states, explicitly they are as follows
%%%%%%%%%%%%%%%%%%%%%%%%%%%%
\begin{align}
\tilde{\Psi}_{++}^{\mu}(p)&=\epsilon_{+}^{\mu}(p)u_+(p),\label{eq:gravitinostate01}\\
\tilde{\Psi}_{--}^{\mu}(p)&=\epsilon_{-}^{\mu}(p)u_-(p),\label{eq:gravitinostate02}\\
\tilde{\Psi}_{+}^{\mu}(p)&=\sqrt{\frac{2}{3}}\epsilon_{0}^{\mu}(p)u_+(p)+\frac{1}{\sqrt{3}}\epsilon_{+}^{\mu}(p)u_-(p),\label{eq:gravitinostate03}\\
\tilde{\Psi}_{-}^{\mu}(p)&=\sqrt{\frac{2}{3}}\epsilon_{0}^{\mu}(p)u_-(p)+\frac{1}{\sqrt{3}}\epsilon_{-}^{\mu}(p)u_+(p).\label{eq:gravitinostate04}
\end{align}
Replacing the momentum twistors in the four gravitino states Eqs.~(\ref{eq:gravitinostate01})-(\ref{eq:gravitinostate04}) is straightforward (See Ref.~\cite{lorenzo-bryan} for more details). Dealing with HAs involving gravitinos in the final state expressed now in these new variables shall avoid the large and messy expressions that appear when the trace technology is used (most of the time handled with \verb|Mathematica|), we have to remember that the completeness relation for the gravitino takes the form
%%%%%%%%%%%%%%%%%%%
 \begin{align}\label{eq:a3:completeness}\nonumber
D_{\mu\nu}(p)=\sum_{\tilde{\lambda}=1}^3\tilde{\Psi}_{\mu}(\vec{p},\tilde{\lambda})\overline{\tilde{\Psi}}_{\nu}(\vec{p},\tilde{\lambda}) &=-(\s{p}+\tilde{m})\times
\left[\left(g_{\mu\nu}-\frac{p_{\mu}p_{\nu}}{\tilde{m}^2}\right)\right. \\
&\quad \left.-\frac{1}{3}\Bigg(g_{\mu\sigma}-\frac{p_{\mu}p_{\sigma}}{\tilde{m}^2}\Bigg)\Bigg(g_{\nu\lambda}-\frac{p_{\nu}p_{\lambda}}{\tilde{m}^2}\Bigg)\g^{\sigma}\g^{\lambda} \right].
\end{align}
Fortunately in the SHF the fundamental building block are the HAs (without  square modulus) and we shall not need Eq.~(\ref{eq:a3:completeness}), neither another completeness relation in general.
%%%%%%%%%%%%%%%%%%% 
\subsection{High energy equivalence theorem}
We briefly discuss in this section the gravitino approximation to goldstino. When the gravitino mass ($\tilde{m}$) is small compared to the energy of the process, it is possible to applicate the equivalence theorem which roughly speaking allow us to replace the longitudinal components of the gravitino by the derivative of the goldstino field \cite{equivalence-theorem, fayet1, fayet2, fayet3, fayet4, fayet5}. In practice the equivalence theorem tell us that the four gravitino states Eqs.~(\ref{eq:gravitinostate01})-(\ref{eq:gravitinostate04}), go as follows; $\tilde{\Psi}_{++}^{\mu}(p)\approx0$, $\tilde{\Psi}_{--}^{\mu}(p)\approx0$, $\tilde{\Psi}_{-}^{\mu}(p)\approx\sqrt{\frac{2}{3}}\left(\frac{p^{\mu}}{\tilde{m}}\right)u_-(p)$ and $\tilde{\Psi}_{+}^{\mu}(p)\approx\sqrt{\frac{2}{3}}\left(\frac{p^{\mu}}{\tilde{m}}\right)u_+(p)$. At the end just two gravitino states shall survive in this approximation, one advantage  is that they just depend on spinors and not anymore on the polarization vectors that makes calculations worse.  
%%%%%%%%%%%%%%%%%%%%%%%%%%%%%%%%%%%%%%%%%%
\section{Decay Amplitudes with LSP gravitino/goldstino }\label{section3}
In this section the idea is to present how the SHF works, with the amplitudes at hands the next step is to write the spinors and polarization vectors (for the gravitino) as momentum twistors, then the SHF makes its magic. We do not need to worry about addressing the massive case, because a massive momentum twistor is expressed in terms of two massless one \cite{spinorsextras,lorenzo-bryan}. %In Appendix \ref{} some basics rules for the momentum twistors has been put.
In this proceeding contribution two decay widths are worked out ($\tilde{\chi}_0\to V \tilde{\psi}^{\mu}$, $V=\gamma, Z$), both within the local supersymmetric extension of the Standard Model with gravitino LSP in the final state and with the neutralino as the NLSP. These calculations have  been already evaluated in References~\cite{covi, ellis6, feng3} using the Feynman approach with the trace technology. 
%%%%%%%%%%%%%%%%%%%%%%%%%%%%%%%%%%%%%%%
%%%%%%%%%%%%%%%%%%%%%%%%%%%%%%%%%%%%%%%
\subsection{Two-body Neutralino decay $\tilde{\chi}_0\to \tilde{\Psi}^{\mu}\,\gamma$}
We build out the amplitudes for the two-body neutralino decays applying the Feynman rules of Ref.~\cite{takeo} for the spin-3/2 gravitino interactions. We shall start with the decay $\tilde{\chi}_0\to \tilde{\Psi}^{\mu}\,\gamma$ corresponding to the diagram of Figure~\ref{fig:neutralinodecay}. The  $\tilde{\chi}_0(p_1)$ denotes the neutralino with momentum $p_1$ and $\tilde{\Psi}^{\mu}(p_2)$ represents the spin-3/2 gravitino (in the goldstino approximation will be $\tilde{G}(p_2)$), and $\gamma(p_3)$ denotes the photon with momentum $p_3$. The amplitude reads as follows 
\begin{align}\label{eq:100}
\mathcal{M}_{\lambda_1,\lambda_2,\lambda_3}&=\frac{C_{\chi\gamma}}{4M}\bar{\Psi}_{\mu\lambda_2}(p_2)(p_{3\nu}[\gamma^{\nu},\gamma^{\sigma}]\gamma^{\mu})\epsilon_{\sigma\lambda_3}(p_3)u_{\lambda_1}(p_1).
\end{align}
where $C_{\chi\gamma}=U_{i1}\cos\theta_W+U_{i2}\sin\theta_W$, $U_{ij}$ are the mixing matrices that diagonalize the neutralino factor, $M$ is the Plank mass, then $\lambda_1$, $\lambda_2$ and $\lambda_3$ are the helicity labels corresponding to the neutralino, gravitino and photon. After applying the SHF to the  amplitude of Eq.~(\ref{eq:100}), one notice that  there are in principle 16 HAs, but 14 of them vanish. The SHF helps to identify the spurious quantities from the beginning, making the process to find physical observables expeditious.  The nonzero HAs are shown in the Table \ref{table1}.
%%%%%%%%%%%%%%%%% table for the neutralino 1%%%%%%%%%%
\begin{table}[H]
\begin{center}
  \begin{tabular}{  || c | c || }
    \hline \hline
    $\lambda_1,\,\lambda_2,\,\lambda_3$ & $\mathcal{M}_{\lambda_1,\,\lambda_2,\,\lambda_3}^{3/2}$  
    \\ \hline
   \hline\hline
     $-,++,+$ & $ \frac{C_{\chi\gamma}[ r_2q_2]^2}{M\langle r_1r_2\rangle}m_{\tilde{\chi}_0}$\\ \hline
     $-,-,-$ & $\frac{C_{\chi\gamma}s_{r_2q_2}}{\sqrt{3}M\tilde{m}[r_2q_2]}\langle r_2q_2\rangle[r_2r_1]$\\ \hline
\hline
    \hline
  \end{tabular}
    \caption{Helicity Amplitudes for the two-body neutralino decay $\tilde{\chi}_{0}\to\tilde{\Psi}^{\mu}\,\gamma$ with LSP gravitino in the final state. We are using  $s_{ij}=-(p_i+p_j)^2$ for the Mandesltam variable. There are two more HAs ($+,--,+$ and $-,-,-$), but they are  the complex conjugate of the two shown in this table, the same criterium shall apply to the next tables.}
  \label{table1}
 \end{center}
\end{table}
%%%%%%%%%%%%%%%%%%%%%%%%%%%%%%%%%%%
%\newpage

%%%%%%%%%%%%%%%%%%%%%%%%%%%%%%%%%%%
The squared and averaged amplitude takes the form
\begin{align}
\langle|\mathcal{M}|^2\rangle&=\frac{C_{\chi\gamma}^2}{2M^2}\big(|\mathcal{M}_{-,++,+}|^2+|\mathcal{M}_{+,--,-}|^2+|\mathcal{M}_{-,-,-}|^2+\mathcal{M}_{+,+,+}|^2\big)\\
&=\frac{C_{\chi\gamma}^2}{2M^2}\left(2\frac{s_{r_2q_2}^2m_{\tilde{\chi}_0}^2}{s_{r_1r_2}}+2\frac{s_{r_2q_2}^2}{3\tilde{m}^2}s_{r_1r_2}\right)\\
&=\frac{C_{\chi\gamma}^2}{M^2}\left(\frac{(m_{\tilde{\chi}_0}^2-\tilde{m}^2)^2}{3\tilde{m}^2}(3\tilde{m}^2+m_{\tilde{\chi}_0}^2)\right)\\
&=\frac{C_{\chi\gamma}^2m_{\tilde{\chi}_0}^6}{M^2}\left(1-\frac{\tilde{m}^2}{m_{\tilde{\chi}_0}^2}\right)^2\left(\frac{1}{3}+\frac{\tilde{m}^2}{m_{\tilde{\chi}_0}^2}\right).\label{eq:101}
\end{align}
%%%%%%%%%%%%%%%%%%%%%%%%%%%%%%%%%
%%%%%%%%%%%%%%%%%% plot 1 %%%%%%%%%%%%%%%%
\vspace{-0.01\linewidth}
\begin{minipage}{\linewidth}
\begin{figure}[H]
\centering
\begin{picture}(0,85)% width and height of the picture
\put(-70,0){\includegraphics[scale=0.6]{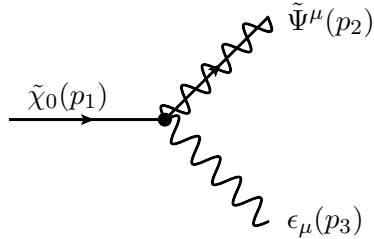}}
\put(-63,47){$\tilde{\chi}_0(p_1)$}
\put(34,0){$\epsilon_{\mu}(p_3)$}
\put(34, 75){$\tilde{\Psi}^{\mu}(p_2)$}
%\put(-20, 25){$b(p_3)$}
%\put(-42,24){$t(l)$}
%\put(70,10){$\Psi(q)$}
%\put(140,52){$V(\tilde{t}\Psi t)=\left(\frac{m_{t}^2-m_{\tilde{t}}^2}{\sqrt{3}M\tilde{m}}\right)(\cos\theta\ \bold{P_R}+\sin\theta\ \bold{P_L})$}
\end{picture}
\caption{Feynman diagram for the 2-body neutralino decay.}
        \label{fig:neutralinodecay}
\end{figure}
      \end{minipage}
     %%% left this space empty
     
       \vspace{0.03\linewidth}

       %%%%%%%%%%%%%%%%%%%%%%%%%%%%%%%%%%%%%%%
The  result of Eq.~(\ref{eq:101})  conduces to the right  decay width found in Ref.~\cite{ellis6, feng3}, but our calculations result considerably simpler applying the SHF (massive).\newline
We shall considerate now the  approximation of the spin-3/2 gravitino to spin-1/2 goldstino, this is due to the  high energy equivalence theorem,  physically the approximation is valid when the split mass between fermions and bosons is larger than the gravitino mass. Some work has been done in order to compare observables (lifetime $\tau$) with gravitino and goldstino in the final state \cite{lorenzo-bryan2}, basically comparing the squared amplitudes. It shall be interesting to exploit the power of the SHF for comparing the HAs with gravitino and goldstino for a given process or reaction.

%%%%%%%%%%%%%%%%%%%%%%%%%%%%%%%%%
%\subsection{2-body Neutralino decay with light gravitino (goldstino) in the final state}

The  amplitude for the decay $\tilde{\chi}_0\to\tilde{G}\,\gamma$ with goldstino in the final state is the following
\begin{equation}
\mathcal{M}_{\lambda_1,\lambda_2,\lambda_3}=\frac{C_{\chi\gamma}m_{\tilde{\chi}_0}}{2\sqrt{6}M\tilde{m}}\bar{\Psi}_{\lambda_2}(p_2)(p_{3\nu}[\gamma^{\nu},\gamma^{\sigma}])\epsilon_{\sigma\lambda_3}(p_3)u_{\lambda_1}(p_1),
\end{equation}
the kinematics of the process remains the same in the   approximation to goldstino, but now the transversal degrees of freedom vanish. The nonzero HAs are shown in the Table \ref{table2}

%what it is reduce is the helicity states for the light gravitino (goldstino), now we just have 2 of the initial 4, these are $\lambda_2=+,-$. There are 8 HAs but just 2 are nonzero, they are shown in the next table
%%%%%%%%%%% 2 body neutralino decay (goldstino) 1 %%%%%%%%%%
  \begin{table}[H]
\begin{center}
  \begin{tabular}{  || c | c || }
    \hline \hline
    $\lambda_1,\,\lambda_2,\,\lambda_3$ & $\mathcal{M}_{\lambda_1,\,\lambda_2,\,\lambda_3}^{1/2}$ \\ \hline
   \hline\hline
    $-,-,-$ & $\frac{C_{\chi\gamma}s_{r_2q_2}}{\sqrt{3}M\tilde{m}[r_2q_2]}\langle r_2q_2\rangle[r_2r_1]$\\ \hline
\hline
    \hline
  \end{tabular}
    \caption{
    Helicity Amplitudes for the two-body neutralino decay 
    $\tilde{\chi}_{0}\to\tilde{G}\,\gamma$ with LSP goldstino in the final state. }%There is one more helicity amplitude ($+,+,+$), but as in Table \ref{table1}, it is just the complex conjugate of the helicity amplitude $\mathcal{M}_{+,+,+}^{1/2}$.
    %}
  \label{table2}
 \end{center}
\end{table}
In Table \ref{table1} we show two HAs, one of them includes the transversal d.o.f. ($-,++,+$) corresponding to the full gravitino contribution, but  Table \ref{table2} shows just one HA, $\mathcal{M}_{-,-,-}^{1/2}$, that is in fact identically to the  HA $\mathcal{M}_{-,-,-}^{3/2}$. %CERRAR COMENTARIO
%%%%%%%%%%%%%%%%%%%%%%%%%%%%%%%%%%%%%%%
\newline
The squared and averaged amplitude takes the following form
\begin{align}
\langle|\mathcal{M}|^2\rangle&=\frac{C_{\chi\gamma}^2}{2M^2}(|\mathcal{M}_{-,-,-}|^2+|\mathcal{M}_{+,+,+}|^2)\\&=\frac{C_{\chi\gamma}^2s_{r_2q_2}^2}{3M^2\tilde{m}^2}s_{r_2r_1}\\
&=\frac{C_{\chi\gamma}^2(m_{\tilde{\chi}_0}^2-\tilde{m}^2)^2m_{\tilde{\chi}_0}^2}{3M^2\tilde{m}^2},
\end{align}
this result can be obtained directly from Eq.~(\ref{eq:101}) in the limit that the gravitino mass ($\tilde{m}$) is small compared to the neutralino mass ($m_{\tilde{\chi}_0}$).
%%%%%%%%%%%%%%%%%%%%%%%%%%%%%%%%%%%%%%%%
%%  body neutralino decay to Z a photon with gravitino full
%%%%%%%%%%%%%%%%%%%%%%%%%%%%%%%%%%%%%%%%%%%%

\subsection{2-body neutralino decay $\tilde{\chi}_0\to \tilde{\Psi}^{\mu}\,Z $}
In the case that the vector boson is massive (but neutral) the decay becomes $\tilde{\chi}_0(p_1)\to \tilde{\Psi}^{\mu}(p_2)\,Z(p_3) $. Some technical complications appear from having the $Z$ boson in the final state instead of the photon.
The amplitude for this process with the spin-3/2 gravitino in the final state is as follows
\begin{equation}\label{eq:201}
\mathcal{M}_{\lambda_1,\lambda_2,\lambda_3}=\frac{C_{\chi Z}}{4M}\bar{\Psi}_{\mu\lambda_2}(p_2)(p_{3}^{\rho}[\gamma_{\rho},\gamma_{\sigma}]\gamma^{\mu})\epsilon_{\lambda_3}^{\sigma}(p_3)u_{\lambda_1}(p_1),
\end{equation}
where $C_{\chi\,Z}=-U_{i1}\sin\theta_W+U_{i2}\cos\theta_W$, and everything else remains equal to the decay with a photon in final state ($\tilde{\chi}_0\to \tilde{\Psi}^{\mu}\,\gamma$), but now the helicity label corresponding to the $Z$ boson ($\lambda_3$) could be $+,-,0$. 

From the 24 possible HAs resulting of Eq.~(\ref{eq:201}), the nonzero are shown in the Table \ref{table3}.
%%%%%%%%% Table 1 %%%%%%%%%%%%%%%%%%%
\begin{table}[H]
\begin{center}
  \begin{tabular}{  || c | c || }
    \hline \hline
    $\lambda_1,\,\lambda_2,\,\lambda_3$ & $\mathcal{M}_{\lambda_1,\,\lambda_2,\,\lambda_3}^{3/2}$ \\ \hline
   \hline\hline
     $+,-,+$ & $\frac{C_{\chi Z}}{\sqrt{3}M\tilde{m}s_{r_2q_2}[q_2r_2][r_1q_2]}\left(\tilde{m}^3M_Z^2s_{q_2r_1}+s_{r_2q_2}^3m_{\tilde{\chi}_0}\right) $\\ \hline
     $-,-,0$ & $\frac{\sqrt{2}C_{\chi Z}M_Z\langle r_2q_2\rangle}{\sqrt{3}Ms_{r_2q_2}\langle r_1q_2\rangle}\left(\tilde{m}s_{q_2r_1}+s_{r_2q_2}m_{\tilde{\chi}_0}\right) $\\ \hline

     $-,--,+$ & $-\frac{C_{\chi Z}\langle r_2q_2\rangle}{M\langle r_1q_2\rangle[r_2q_2]^2}(\tilde{m}M_Z^2m_{\tilde{\chi}_0}+s_{r_2q_2}s_{q_2r_1})$\\ \hline
\hline
    \hline
  \end{tabular}
    \caption{Helicity Amplitudes for the two-body Neutralino decay $\tilde{\chi}_{0}\to\tilde{\Psi}^{\mu}\, Z$ with LSP gravitino in the final state.}
  \label{table3}
 \end{center}
\end{table}

%%%%%%%%%%%%%%%%%%%%%%%%%%%%%%%%%%
The squared and averaged amplitude is as follows
\begin{align}
\langle|\mathcal{M}|^2\rangle&=\frac{C_{\chi Z}^2}{2M^2}(|\mathcal{M_{+,-,+}}|^2+|\mathcal{M_{-,+,-}}|^2+|\mathcal{M}_{-,-,0}|^2+|\mathcal{M}_{+,+,0}|^2+|\mathcal{M}_{-,--,+}|^2+|\mathcal{M}_{+,++,-}|^2)\\\nonumber
&=\frac{C_{\chi Z}^2}{M^2}\Bigg[\frac{1}{3\tilde{m}^2s_{r_2q_2}^3s_{r_1q_2}}\left(\tilde{m}^3M_Z^2s_{q_2r_1}+s_{r_2q_2}^3m_{\tilde{\chi}_0}\right)^2+\frac{2M_Z^2}{3s_{r_2q_2}s_{q_2r_1}}(s_{r_2q_2}m_{\tilde{\chi}_0}+\tilde{m}s_{q_2r_1})^2\\ 
&\quad
+\frac{1}{s_{q_2r_1}s_{r_2q_2}}(\tilde{m}m_{\tilde{\chi}_0}M_{Z}^2+s_{q_2r_1}s_{r_2q_2})^2
\Bigg]\\\nonumber
&=\frac{C_{\chi Z}^2m_{\tilde{\chi}_0}^6}{M^2\tilde{m}^2}\Bigg[\Bigg(1-\frac{\tilde{m}^2}{m_{\tilde{\chi}_0}^2}\Bigg)^2\left(\frac{1}{3}+\frac{\tilde{m}^2}{m_{\tilde{\chi}_0}^2}\right)-\frac{M_Z^2}{m_{\tilde{\chi}_0}^2}\Bigg(1-\frac{M_Z^2}{m_{\tilde{\chi}_0}^2}\left(1-\frac{\tilde{m}^2}{3m_{\tilde{\chi}_0}^2}\right)-\frac{\tilde{m}^3}{m_{\tilde{\chi}_0}^3}\left(4-\frac{\tilde{m}}{3m_{\tilde{\chi}_0}}\right)
\\
&\quad
+\frac{M_Z^4}{3m_{\tilde{\chi}_0}^4}\Bigg)\Bigg],
\end{align}
%%%%%%
this result conduces to the known  decay width  found in \cite{ellis6, feng3}.
%%%%%%%%%%%%%%%%%%%%%%%%%%%%%%%%%%%%%%%
%%%%%% 2 body neutralino decay with goldstino
%%%%%%%%%%%%%%%%%%%%%%%%%%%%%%%%%%%%%%%
%\subsection{2-body neutralino decay with light gravitino in the final state}
\newline
We repeat the process of the first example and approximate the gravitino to goldstino, then we compute the HAs.
\begin{equation}
\mathcal{M}_{\lambda_1,\lambda_2,\lambda_3}=\frac{C_{\chi Z}m_{\tilde{\chi}_0}}{2\sqrt{6}M\tilde{m}}\bar{\Psi}_{\lambda_2}(p_2)(p_3^{\rho}[\gamma_{\rho},\gamma_{\sigma}])\epsilon_{\lambda_3}^{\sigma}(p_3)u_{\lambda_1}(p_1),
\end{equation}
the nonzero HA are shown in the  Table \ref{table4}
%%%%%%%%%%%%%%%% table 2 %%%%%%%%%%%
\begin{table}[H]
\begin{center}
  \begin{tabular}{  || c | c || }
    \hline \hline
    $\lambda_1,\,\lambda_2,\,\lambda_3$ & $\mathcal{M}_{\lambda_1,\,\lambda_2,\,\lambda_3}^{1/2}$ \\ \hline
   \hline\hline
     $+,-,+$ & $\frac{C_{\chi Z}m_{\tilde{\chi}_0}}{\sqrt{3}M\tilde{m}[q_2r_2][r_1q_2]}\left(\tilde{m}m_{\tilde{\chi}_0}M_Z^2+s_{r_2q_2}s_{q_2r_1}\right) $\\ \hline
     $-,-,0$ & $\frac{C_{\chi Z}M_Zm_{\tilde{\chi}_0}\langle r_2q_2\rangle}{\sqrt{3}M\tilde{m}s_{r_2q_2}\langle r_1q_2\rangle}\left(\tilde{m}s_{q_2r_1}+s_{r_2q_2}m_{\tilde{\chi}_0}\right) $\\ \hline
\hline
    \hline
  \end{tabular}
    \caption{Helicity Amplitudes for the two-body Neutralino decay $\chi_{0}\to Z\,\tilde{G}$ with goldstino in the final state.}
  \label{table4}
 \end{center}
\end{table}
%%%%%%%%%%%%%%%%%%%%%%%%%%%%%%%%%%%%%%%%%%%%%%%%%%%%%%%%%%%%%% 
We can notice in Table \ref{table3} that for the helicity labels $\lambda_1=-$, $\lambda_2=-$ and $\lambda_3=0$ the HA for the gravitino is up to a factor of $\sqrt{2}$ the same HA with  goldstino, this is $\mathcal{M}_{-,-,0}^{3/2}=\sqrt{2}\mathcal{M}_{-,-,0}^{1/2}$.
Squaring and averaging the HAs of Table \ref{table4}, these take the following form
\begin{align}
\langle |\mathcal{M}|^2\rangle&=(|\mathcal{M}_{+,-,+}|^2+|\mathcal{M}_{-,+,+}|^2+|\mathcal{M}_{-,-,0}|^2+|\mathcal{M}_{+,+,0}|^2)\\
&=\frac{C_{\chi Z}^2m_{\tilde{\chi}_0}^2}{M^2\tilde{m}^2s_{r_2q_2}s_{q_2r_1}}\left(M_Z^2(m_{G}s_{q_2r_1}+s_{r_2q_2}m_{\tilde{\chi}_0})^2+2(\tilde{m}m_{\tilde{\chi}_0}M_Z^2+s_{r_2q_2}s_{q_2r_1})^2\right)\\
&=\frac{2C_{\chi Z}^2m_{\tilde{\chi}_0}^6}{M^2\tilde{m}^2}\left[\left(1-\frac{\tilde{m}^2}{m_{\tilde{\chi}_0}^2}\right)^2-\frac{M_Z^2}{m_{\tilde{\chi}_0}^2}\left(\frac{1}{2}\left(1-6\frac{\tilde{m}}{m_{\tilde{\chi}_0}}+\frac{\tilde{m}^2}{m_{\tilde{\chi}_0}^2}\right)+\frac{M_Z^2}{2m_{\tilde{\chi}_0}^2}\right)\right]
\end{align}
In the traditional Feynman approach with trace technology, the calculations of the squared and averaged amplitudes with goldstinos are considerably simple, unlike amplitudes involving gravitinos. However, when the SHF is implemented the calculations with gravitino are still simple, and the helicity method is able to compute the decay widths without any approximation, as we have shown.

%\begin{figure}[H]
%	\begin{center}
%		\includegraphics[scale=0.5]{h_decay_zff.png}
%	\end{center}
%	\label{w_decay}
%	\caption{Higgs decay into a $Z$ and a fermion-antifermion pair.}
%\end{figure}
%%%%%% Conclussions %%%%%%%%%%%%%%%%%%%%%%%
\section{Conclusions}\label{conclussions}

In this proceeding, we have presented how the spinor helicity formalism is well suitable to evaluate decay amplitudes, even in the massive case. Two examples have been recalculated with the SHF ($\tilde{\chi}_0\to V\tilde{\Psi}^{\mu}$ with  $V=\gamma, Z$). Furthermore, we briefly discussed the high energy equivalence theorem between gravitinos and goldstinos. It is possible in some cases to identify how the helicity amplitudes with goldstinos appear as a helicity amplitudes with gravitinos (for the longitudinal d.o.f.), this characteristic is dificult to identify in the traditional approach.  We shall study and discuss more relationships between gravitino and goldstino amplitudes for another NLSP candidates in a future work to appear soon in a peer review journal.
%we shown how in some cases the helicity amplitudes with goldstinos in the final state appear directly in the helicity amplitudes with gravitinos, this is characteristic shall not be always obvious to observe.   

%has been applied to efficiently compute  perturbative scattering amplitudes for a plenty of processes and reactions in gauge theories (including gravity), mostly in the massless case. Some work has been done in order to extend the SHF to the massive case. We have used this powerful tools to evaluate amplitudes in a local supersymmetric model where the gravitino is the lightest supersymmetric particle (LSP).  Two decays has been evaluated in order to shown the capability of the SHF in the massive extentiion, namely the two body neutralino decays $\tilde{\chi}_0\to V\tilde{\Psi}^{\mu}$ with $V=\gamma, Z$, the comparisons of amplitudes with spin-3/2 gravitino and spin-1/2 goldstino are also presented.

% we have presented a summary of the basic formulae of the SHF.  In order to appreciate the value of the methods, we studied the phenomenology of the Electroweak sector of the Standard Model of Particle Physics, including the evaluation of the decays: $Z\to ff$, $h\to ff$, $h\to W^{-}W^{+}$ and $h\to V f'\bar{f}$. 
%Although these results are well known, it can be appreciated that the simplification obtained by using SHF, make it worth to use them in teaching of Particle Physics within a modern approach.
%%%%%%%%%%%%%%%%%%%%%%%%%%%%%%%%%%%%%%%
\section{Acknowledgements}
The author wish to thank  Lorenzo Diaz-Cruz for useful discussion, Roberto Garcia and Jose Zelaya for reading the manuscript. This work has been supported by CONACYT  and was  partly supported by DPC-SMF.
%%%%%%%%%%%%%%%%%%%%%%%%%%%%%%%%%%%%%%%
%%%%%%%%%%%%%%%%%%%%% References %%%%%%%%%%


\section*{References}
\begin{thebibliography}{99}
\bibitem{dixon} L.~Dixon, arXiv:hep-ph/9601359.

\bibitem{parke} 
S.J. Parke and T. Taylor, Phys. Lett. 157 B:81 (1985); Z. Kunszt, Nucl. Phys. B 271:333 (1986);
M. Mangano and S. Parke, Phys. Rep. 200:301 (1991).

%\bibitem{iopartnum} IOP Publishing is to grateful Mark A Caprio, Center for Theoretical Physics, Yale University, for permission to include the {\tt iopart-num} \BibTeX package (version 2.0, December 21, 2006) with  this documentation. Updates and new releases of {\tt iopart-num} can be found on \verb"www.ctan.org" (CTAN). 
%%%%%% ref 1 %%%%%%
\bibitem{srednicki} M. Srednicki, \textit{Quantum Field Theory}, Cambridge University Press, 2007.
%%%%%% ref 2 %%%%%%
%%%%%% ref 5 %%%%%%
\bibitem{peskin2} M.~Peskin, http://arxiv.org/pdf/1101.2414.pdf.
%%%%%% ref 3 %%%%%%
\bibitem{schwartz} M. D. Schwarz, \textit{Quantum Field Theory and the Standard Model}, Cambridge University Press, 2014.
%%%%%%%%%%%%%%%%

\bibitem{henn}
J.~M.~Henn, J.~C. Plefka, \textit{Scattering Amplitudes in Gauge Theories}, in: Lecture Notes in Physics, vol. 883, Springer, 2014.
%%%%%%%%%%%%%%
\bibitem{elvang} H. Elvang and Y. Huang, \textit{Scattering Amplitudes in Gauge Theory and Gravity} (Cambridge University Press, 2015), arXiv:1308.1697v2 [hep-th].
%%%%%%%%%%%%%%%%


%%%%%%%% ref 6%%%%%%%%
%%%%%%%%%%%%%%%%

%\bibitem{mangano-parke} F.~A.~Berends and W.~T.~Giele, Nucl.Phys.B294:700(1987);
%M.~Mangano, S.~Parke and Z.~Xu, Nucl. Phys. B298:653 (1988); 
%M. Mangano, Nucl.
%Phys. B309:461 (1988)
%%%%%% ref 4 %%%%%%
%\bibitem{bern}
%Z.~Bern, L.~Dixon and D.~A.~Kosower, Nucl. Phys. B 437:259 (1995).

%%%%%%%%%%%%%%%%%%%%%%%
\bibitem{witten} E. Witten, Commun. Math. Phys. 252 189-258 (2004), arXiv:hep-th/0312171.

%%%%%%%%%%%%%%%%%%%%%%%
\bibitem{bcfw}
R. Britto, F. Cachazo, B. Feng, E. Witten, Phys. Rev. Lett. 94 (2005) 181602. hep-th/0501052.
%%%%%%%%%%%%%%%%%%%
\bibitem{mazon}
L.J. Mason, D. Skinner, Phys. Lett. B 636 (2006) 60. hep-th/0510262.
%%%%%%%%%%%%%%%%%%%%%%%
\bibitem{carrasco}
Z.~Bern, J.~J.~M.~Carrasco, H.~Johansson, Phys. Rev. D 78 (2008) 085011. arXiv:0805.3993.
%%%%%%%%%%%%%%%%%%%%%%%

%%%%%%%%%%%%%%%%%%%%%%
%\bibitem{nima1} 
%N. Arkani-Hamed, F. Cachazo, C. Cheung, J. Kaplan, J. High Energy Phys. 03 (2010) 110. arXiv:0903.2110.
%\bibitem{nima2} 
%N. Arkani-Hamed, F. Cachazo, C. Cheung, J. Kaplan, J. High Energy Phys. 03 (2010) 020. arXiv:0907.5418.
%\bibitem{nima3} 
%N. Arkani-Hamed, J. Bourjaily, F. Cachazo, J. Trnka, J. High Energy Phys. 01 (2011) 108. arXiv:0912.3249.
%\bibitem{nima4} 
 %N. Arkani-Hamed, J. Bourjaily, F. Cachazo, J. Trnka, J. High Energy Phys. 01 (2011) 049. arXiv:0912.4912.
\bibitem{nima5} 
N. Arkani-Hamed, J. Trnka, J. High Energy Phys. 10 (2014) 030. arXiv:1312.2007.
%\bibitem{nima6} 
%N. Arkani-Hamed, J. Trnka, J. High Energy Phys. 12 (2014) 182. arXiv:1312.7878.
%\bibitem{nima7} N. Arkani-Hamed, A. Hodges, J. Trnka, J. High Energy Phys. 08 (2015) 030. arXiv:1412.8478.
\bibitem{cachazo}
F. Cachazo, S. He, E.Y. Yuan, Phys. Rev. D 90 (2014) 065001. arXiv:1306.6575.
%\bibitem{cachazo2} 
%F. Cachazo, S. He, E.Y. Yuan, Phys. Rev. Lett. 113 (2014) 171601. arXiv:1307.2199.
%\bibitem{cachazo3}
 %F. Cachazo, S. He, E.Y. Yuan, J. High Energy Phys. 1407 (2014) 033. arXiv:1309.0885.
%\bibitem{cachazo4}
%L. Dolan, P. Goddard, J. High Energy Phys. 1407 (2014) 029. arXiv:1402.7374.

%%%%%%%%%%%%%%%
%\bibitem{dittmaier} S. Dittmaier, Phys. Rev. D 59 016007 (1998), hep-ph/9805445.
%%%%%% ref 7 %%%%%%
%\bibitem{badger1} S. Badger, E. N. Glover, V. Khoze, and P. Svrcek, JHEP \textbf{0507} 025 (2005), arXiv:hep-th/0504159.
%%%%%% ref 8 %%%%%%
%\bibitem{badger2} S. Badger, E. N. Glover, and V. V. Khoze, JHEP \textbf{0601} 066 (2006), arXiv:hep-th/0507161.
%%%%%% ref 9 %%%%%%
\bibitem{ozeren} K. Ozeren and W. Stirling, Eur. Phys. J. C 48 159 (2006), arXiv:hep-ph/0603071.
%%%%%%%% ref 10 %%%
%\bibitem{hall} A. Hall, Phys.Rev. D 77 025011 (2008), arXiv:0710.1300 [hep-ph].
%%%%%% ref 11 %%%%%%
%\bibitem{boels} R. Boels, JHEP 1001 010 (2010), arXiv:0908.0738 [hep-th].
%%%%%% ref 12 %%%%%%
%\bibitem{dixon2} L. J. Dixon, J. M. Drummond, C. Duhr and J. Pennington, JHEP 1406 116 (2014), arXiv:1402.3300 [hep-th].
%%%%%% ref 13 %%%%%%
%\bibitem{bcfw} R. Britto, F. Cachazo, B. Feng, and E. Witten, %Direct proof of tree-level recursion relation in Yang-Mills theory, 
%Phys. Rev. Lett. 94 (2005) 181602, [hep-th/0501052].
%%%%%%%%%%%%%%%%%
%\bibitem{sam} D. Ma\^itre and P. Mastrolia, Comput. Phys. Commun. 179 501-574 (2008), arXiv:0710.5559 [hep-ph].
%%%%%% ref 14 %%%%%%
\bibitem{spinorsextras} J. Kuczmarski, arXiv:1406.5612 [hep-ph].
%%%%%% ref 15 %%%%%%
%\bibitem{rarita} W. Rarita and J. Schwinger, Phys. Rev. \textbf{60} (1941) 60.
%%%%%% ref 15 %%%%%%
%\bibitem{novaes} S. F. Novaes and D. Spehler, Nucl. Phys. B \textbf{371} (1992) 618-636.
%%%%%% ref 15 %%%%%%
%\bibitem{us} Bryan O.~Larios, O. Meza, Work in progress.
%%%%%%% ref 16 %%%%%%
\bibitem{kosower} D. A. Kosower, Phys. Rev. D 71, 045007 (2005), [arXiv:hep-th/0406175].
%\bibitem{takeo} T. Moroi, arXiv:hep-ph/9503210.
%\bibitem{ryder} L. H. Ryder, \textit{Quantum Field Theory} (Cambridge University Press, 1996).
%%%%%% ref 16 %%%%%%
%\bibitem{peskin} M. E. Peskin and D. V. Schroeder, \textit{An Introduction to Quantum Field Theory} (Addison Wesley, 1995).
%%%%%% ref 17 %%%%%%
%\bibitem{lorenzo} J. L. D\'iaz Cruz, B. Larios, O. Meza Aldama and J. Reyes P\'erez, Rev. Mex. Fis. E 61(2) (2015) 104. English version: arXiv:1511.07477 [physics.gen-ph].
%%%%%% ref 18 %%%%%%
\bibitem{weinzierl}
C. Schwinn and S. Weinzierl, JHEP 0704 (2007) 072, arXiv:hep-ph/0703021 [HEP-PH].
%%%%%%%%%%%
\bibitem{bryan-oscar}
J.~Lorenzo Diaz-Cruz, Bryan O. Larios, O. Meza-Aldama, J.~Phys.~Conf.~Ser.~761 (2016) no.1, 012012,  arXiv:1608.04129 [hep-ph].  
%%%%%%%%%%%%%%%%%%%%
\bibitem{lorenzo-bryan}
J. Lorenzo Diaz-Cruz, Bryan O. Larios, arXiv:1612.04331 [hep-ph].
%%%%%%%%%% Bryan references %%%%%%%%
%%%%%% Bryan reference %%%%%
\bibitem{feng1}
J. L. Feng, A. Rajaraman and F. Takayama, Phys. Rev. Lett. 91 (2003) 011302,
arXiv:hep-ph/0302215; Phys. Rev. D 68 (2003) 063504, arXiv:hep-ph/0306024.
%%%%%%%%%%%%%%%%%%%%
 \bibitem{feng2}
J. L. Feng, S. Su and F. Takayama, Phys. Rev. D 70 (2004) 075019, arXiv:hep-ph/0404231.
%%%%%%%%%%%%%%%%%%%%
\bibitem{ellis1}
 J. R. Ellis, K. A. Olive, Y. Santoso and V. C. Spanos, Phys. Lett. B 588 (2004) 7, arXiv:hep-ph/0312262.
%%%%%%%%%%%%%%%%%%%%
\bibitem{ellis2}
J. L. Diaz-Cruz, John Ellis, Keith A. Olive, Yudi Santoso, 	JHEP 0705:003, 2007, arXiv:hep-ph/0701229v1.
%%%%%%%%%%%%%%%%%%%%
\bibitem{lorenzo-bryan2}
J. Lorenzo D\'iaz-Cruz, Bryan O. Larios, 
 Eur. Phys. J. C76 (2016) no.3, 157,
arXiv:1510.01447v2 [hep-ph]. 
 %%%%%%%%%%%%%%%%%
 \bibitem{steffen}
 F. D. Steffen, arXiv:hep-ph/0711.1240.
 %%%%%%%%%%%%%%%%
 \bibitem{johansen}
  M. Johansen, J. Edsj, S. Hellman, J. Milstead , JHEP 1008 1-27 (2010). 
%%%%%%%%%%%%%%%%%
\bibitem{kribs}
G. D. Kribs, A. Martin, and T. S. Roy, JHEP 0901 (2009) 023, arXiv:hep-ph/0807.4936.
%%%%%%%%%%%%%%%%%%%
\bibitem{heisig}
J. Heisig, J. Heising, JCAP 04 (2014) 023, arXiv:1310.6352.
%%%%%%%%%%%%%%%%%%
\bibitem{ellis3}
J. R. Ellis, A. R. Raklev and O. K. Oye, JHEP 0610, 061 (2006), arXiv:hep-ph/0607261.
%%%%%%%%%%%%%%%%%%
\bibitem{kuno}
K. Hamaguchi, Y. Kuno, T. Nakaya and M. M. Nojiri, Phys. Rev. D 70 (2004) 115007, arXiv:hep-ph/0409248.
%%%%%%%%%%%%%%%%%%
\bibitem{ellis5}
R. H. Cyburt, J. R. Ellis, B. D. Fields and K. A. Olive, Phys. Rev. D 67 (2003) 103521, arXiv:astro-ph/0211258; J. R. Ellis, K. A. Olive and E. Vangioni, Phys. Lett. B 619 (2005) 30, arXiv:astro-ph/0503023.
%%%%%%%%%%%%
\bibitem{kohri}
M. Kawasaki, K. Kohri and T. Moroi, Phys. Lett. B 625 (2005) 7, arXiv:astro-ph/0402490; Phys. Rev. D 71 (2005) 083502, arXiv:astro-ph/0408426.
%%%%%%%%%%%
\bibitem{kohri2}
K. Kohri and Y. Santoso, Phys. Rev. D 79, 043514 (2009), arXiv:0811.1119 [hep-ph].
%%%%%%%%%%
\bibitem{rarita}
W. Rarita and J. Schwinger, Phys. Rev. 60, 61 (1941).
%%%%%%%%%
\bibitem{takeo}
Takeo Moroi, arXiv:hep-ph/9503210v1. 
%%%%%%%%%%
\bibitem{auvil}
P.R. Auvil and J.J. Brehm, Phys. Rev. 145 (1966) 1152.
%%%%%%%%%
\bibitem{equivalence-theorem}
R. Casalbuoni, S. De Curtis, D. Dominici, F. Feruglio, and R. Gatto, Physics Letters B 215, 313-316, 1988. 
%%%%%%%%%%%%%%%%%%
\bibitem{fayet1}
P. Fayet, Phys.Lett. 70B (1977) 461.
\bibitem{fayet2}
P. Fayet, Phys. Lett. B. 175 (1986) 471.
\bibitem{fayet3}
P. Fayet, Phys. Lett. B. 84 (1979) 421.
\bibitem{fayet4}
P. Fayet, Phys. Lett. B. 86 (1979) 272.
\bibitem{fayet5}
P. Fayet, Conference Proc. LPTENS-81-9 (1981) 347.
%%%%%%%%%%%%%%%%%%
\bibitem{covi}
Laura Covi, Jasper Hasenkamp, Stefan Pokorski, Jonathan Roberts, JHEP 0911:003, 2009, arXiv:0908.3399v1 [hep-ph].
\bibitem{ellis6}
John Ellis, Keith A. Olive, Yudi Santoso, Vassilis Spanos, 
Phys.~Lett.~B588:7-16, 2004,  arXiv:hep-ph/0312262v4.
\bibitem{feng3}
Jonathan L. Feng, Shufang Su, Fumihiro Takayama, 	Phys. Rev. D70:075019, 2004, arXiv:hep-ph/0404231v2.


\end{thebibliography}
\end{document}